\pdfoutput=1 
\documentclass[review]{elsarticle}
\usepackage{color}
\usepackage{xcolor}
\usepackage{soul}
\usepackage{hyperref}
\hypersetup{
	colorlinks=true,
		citecolor=red,
	linkcolor=blue,
	filecolor=black,      
	urlcolor=red,
}
\usepackage{times}
\usepackage[paper=a4paper,left=25mm,right=25mm,top=20mm,bottom=25mm]{geometry}
\usepackage{graphicx}
\usepackage{epstopdf}
\usepackage{fancyhdr}
\usepackage{fancybox}
\usepackage[]{amsmath}
\usepackage{subfigure}
\usepackage{float}
\usepackage[normalem]{ulem} 
\usepackage{amssymb}
\usepackage{mathrsfs}
\usepackage{tabularx}
\usepackage{booktabs}
\usepackage{bm}
\usepackage{natbib}
\usepackage{amsthm}

\journal{XXX}

\biboptions{square,numbers,sort&compress}

\begin{document}
\begin{frontmatter}
\title{Physical structures of boundary fluxes of orbital rotation and spin for incompressible viscous flow}

\author[1]{Tao Chen\corref{mycorrespondingauthor}}
\cortext[mycorrespondingauthor]{Corresponding author}
\ead{chentao2023@njust.edu.cn}

\address[1]{School of Physics, Nanjing University of Science and Technology, Nanjing 210094, China}

%
%
\begin{abstract}
Vorticity is locally created on a boundary at the rate measured by the boundary vorticity flux, which can be further decomposed as the sum of the orbital rotation and the (generalized) spin.
For incompressible viscous flow interacting with a stationary wall, the full expressions of the boundary fluxes of the orbital rotation and the spin are derived, for the first time, to elucidate their boundary creation mechanisms, respectively. Then, these new findings are successfully extended to the study of the boundary enstrophy dynamics, as well as the Lyman vorticity dynamics as an alternative interpretation to the boundary vorticity dynamics.
Interestingly, it is found that the boundary coupling of the longitudinal and transverse processes is only embodied in the boundary spin flux, which is definitely not responsible for the generation of the boundary orbital-rotation flux.
In addition, the boundary fluxes of enstrophy are directly associated with the boundary source of the second principal invariant of the velocity gradient tensor, and the two quadratic forms representing the spin-geometry interaction.
The present exposition provides a new perspective to the vorticity dynamics on boundaries, which could be valuable in clarifying the formation mechanisms of near-wall coherent structures and flow noise at the fundamental level.
\end{abstract}

\end{frontmatter}
\section{Introduction}
Vorticity (defined as the curl of the velocity field, $\bm{\omega}\equiv\bm{\nabla}\times\bm{u}$) represents the locally-averaged rotation rate of a fluid element, which has been widely used to identify the coherent vortical structures, being viewed as the sinews and muscles of complex fluid motions. Compared to the momentum description constituted by the original flow variables (velocity and pressure), the vorticity description exhibits the potential to provide us with a deeper understanding and a more intuitive picture of coherent flow structures~\citep{Batchelor1967,WuJZ2015book}. 
For incompressible viscous flow, it has been recognized that a solid boundary is the unique source of vorticity~\citep{WuJZ2015book,Lighthill1963}. Therefore, the boundary vorticity dynamics, accounting for the boundary creation mechanisms of vorticity at the fundamental level, was gradually established and became an important research topic in probing the physics of vorticity-boundary interaction.

The discussion on physical measures of the boundary vorticity creation rate was pioneered by Lighthill~\citep{Lighthill1963} who first introduced the concept of the \textit{boundary vorticity flux} (BVF) $\bm{\sigma}\equiv\nu\left[\partial_{n}\bm{\omega}\right]_{w}$ for two-dimensional (2D) incompressible viscous flow interacting with a stationary flat wall (where $\nu$ is the kinematic viscosity, $\bm{n}$ is the unit wall-normal vector directed into the fluid, and the subscript $w$ represents the restriction of a physical quantity to the wall). By applying the Navier-Stokes (NS) equations to the wall and using the velocity adherence condition, the on-wall relation proposed by Lighthill~\citep{Lighthill1963} actually implies that the surface pressure gradient creates the new vorticity at the rate measured by the BVF, by virtue of the viscosity and the no-slip boundary condition. This definition was later generalized by Panton~\citep{Panton1984} to three-dimensional (3D) viscous flows while the full expression of the BVF was derived by Wu et al.~\citep{WuJZWuJM1993,WuJZ2015book} in a set of papers. Rather than using the BVF to measure the boundary creation rate of vorticity, Lyman~\citep{Lyman1990} suggested an alternative definition (namely,~\textit{the Lyman vorticity flux}) as $\bm{\sigma}^{\prime}\equiv-\nu\left[\bm{n}\times\left(\bm{\nabla}\times\bm{\omega}\right)\right]_{w}$ by absorbing the pure 3D viscous contribution to the definition of $\bm{\sigma}$. The integrals of these two BVFs are exactly the same when integrated over a closed surface, although they should be differentiated in describing the local vorticity creation rate. It should be mentioned that the two BVFs correspond to different definitions of the viscid vorticity current tensor, which leads to two different, but equally valid, dynamical interpretations of vorticity and enstrophy dynamics (namely, the Lighthill-Panton and Huggins-Lyman interpretations)~\citep{Terrington2023JFM,Chen2024AMS}. The applications of these two BVFs in certain physical problems can be found in some existing works~\cite{WuJZ2015book,XinWu2013,ChenYu2017}.

The boundary enstrophy flux (BEF) $F_{\Omega}\equiv\nu\left[\partial_{n}\Omega\right]_{w}=\bm{\sigma}\bm{\cdot}\bm{\omega}_{w}$ is directly related to the BVF ($\Omega\equiv\omega^{2}/2$ is the enstrophy), which describes the creation rate of enstrophy on a boundary through a scalar quantity~\citep{WuJZ1995,Liu2016MST,ChenTao2019POF}.
The BEF is an intriguing quantity that is particularly related to the topological features such as isolated critical points and separation/attachment lines in a skin-friction field.
An explicit relation among skin friction, surface pressure gradient and surface curvature tensor has been derived~\citep{Liu2016MST,ChenTao2019POF} and found some new applications~\citep{Liu2019PAS}: by modeling the BEF properly, global skin friction and surface pressure fields with high resolution (at one vector per pixel) can be mutually inferred by virtue of the variational method based on the conventional experimental measurements such as the 
global luminescent oil-film (GLOF) method for skin friction and the pressure sensitive paint (PSP) for surface pressure.
Further exploration focuses on the characteristic surface quantities and exact coupling relations
between near-wall viscous flow and surface physical quantities~\cite{ChenTao2021POF,Chen2023PhysicaD,ChenLiu2023LD}. The BEF was also shown to be associated with the boundary flux of the Lamb vector divergence~\citep{Chen2023PhysicaD}, and the Lie derivatives of the elementary surface quantities with respect to the propagation velocity of fluid-dynamic variables in near-wall viscous flows~\citep{ChenLiu2023LD}. Both the BVF and BEF have also been extended to the investigation of interfacial vorticity dynamics at fluid-fluid interface from both the sharp-interface and diffuse-interface perspectives~\cite{WuJZ1995,Terrington2022JFM,ChenTao2024IJMF}.
In addition, the boundary fluxes of the velocity gradient tensor (VGT) and associated invariants were investigated, which provided a necessary theoretical basis for the Lagrangian VGT dynamics with solid boundaries~\cite{ChenWu2024}.

Consider a regular streamline $\mathcal{C}$ with the velocity $\bm{u}=q\hat{\bm{t}}\left(q\neq{0}\right)$ ($q\equiv\lVert\bm{u}\rVert$ is the velocity magnitude and $\hat{\bm{t}}=\bm{u}/q$ is the unit tangent vector of the streamline), the vorticity can be decomposed with respect to the \textit{Frenet-Serret} intrinsic triad $(\hat{\bm{t}},\hat{\bm{n}},\hat{\bm{b}})$ attached to the streamline ($\hat{\bm{n}}$ is the principal normal vector and $\hat{\bm{b}}\equiv\hat{\bm{t}}\times\hat{\bm{n}}$ is the binormal vector):
\begin{equation}\label{OI_II}
	\bm{\omega}=\bm{\omega}_{\rm{I}}+\bm{\omega}_{\rm{II}},
\end{equation}
where $\bm{\omega}_{\rm{I}}$ and $\bm{\omega}_{\rm{II}}$ are respectively given by~\cite{Serrin1959}
\begin{subequations}\label{omega12}
	\begin{equation}\label{omega1}
		\bm{\omega}_{\rm{I}}\equiv q\bm{\nabla}\times\hat{\bm{t}}=\kappa{q}\hat{\bm{b}}+q\Xi\hat{\bm{t}},~~~\Xi\equiv\hat{\bm{t}}\bm{\cdot}\bm{\nabla}\times\hat{\bm{t}},
	\end{equation}
	\begin{equation}\label{omega2}
		\bm{\omega}_{\rm{II}}\equiv\bm{\nabla}q\times\hat{\bm{t}}=\hat{\bm{n}}\partial_{b}q-\hat{\bm{b}}\partial_{n}q.
	\end{equation}
\end{subequations}
It is observed that $\bm{\omega}_{\rm{I}}$ is proportional to the curl of $\hat{\bm{t}}$: the binormal component $\kappa{q}$ represents the orbital rotation with the curvature radius of $r=\kappa^{-1}$ in the osculating plane (spanned by by $\hat{\bm{t}}$ and $\hat{\bm{n}}$) and the rotation axis being parallel to $\hat{\bm{b}}$ while the tangential component $q\Xi=q(\hat{\bm{b}}\bm{\cdot}\partial_{n}\hat{\bm{t}}-\hat{\bm{n}}\bm{\cdot}\partial_{b}\hat{\bm{t}})$ represents the tangential vorticity component along a streamline owing to the variation of the directions of the neighboring streamlines. $\bm{\omega}_{\rm{II}}$ describes the shearing effects determined by the gradients of the velocity magnitude in the normal plane (spanned by $\hat{\bm{n}}$ and $\hat{\bm{b}}$).
The preceding analysis implies an orientation-dependent intrinsic decomposition of the vorticity as
\begin{equation}\label{ORS}
	\bm{\omega}=2\bm{\psi}+\bm{s}=\bm{R}+\bm{s},
\end{equation}
where~\textit{the orbital rotation} $\bm{\psi}$ and~\textit{the (generalized) spin} $\bm{s}$ are respectively identified as
\begin{subequations}\label{RS}
	\begin{eqnarray}\label{orbital_rotation}
		\bm{\psi}\equiv\kappa{q}\hat{\bm{b}},~~\bm{R}\equiv2\bm{\psi},
	\end{eqnarray}
	\begin{eqnarray}\label{spin}
		\bm{s}\equiv\bm{\omega}_{h}+\bm{\omega}_{\rm{II}}-\bm{\psi}=q\Xi\hat{\bm{t}}+\hat{\bm{n}}\partial_{b}q-\hat{\bm{b}}\left(\partial_{n}q+\kappa{q}\right),
	\end{eqnarray}
\end{subequations}
In Eq.~\eqref{orbital_rotation}, $\bm{R}$ represents the vorticity due to the orbital rotation like that in rigid-body kinematics, being equal to twice of the angular velocity of a fluid particle moving along a curve.
In Eq.~\eqref{spin}, the term $\bm{\omega}_{h}\equiv{q}\Xi\hat{\bm{t}}=q^{-2}\bm{u}h$ is referred to as~\textit{the streaming vorticity}~\footnote{ $h\bm{u}=q^{2}\bm{\omega}_{h}$ is the advection flux vector of helicity, which is proportional to the streaming vorticity $\bm{\omega}_{h}$ describing the advection of the tangential vorticity by the local velocity.}, where $h\equiv\bm{u}\bm{\cdot}\bm{\omega}$ is the helicity density. The present study adopts a generalized spin with the streaming vorticity as a part of its definition~\footnote{In Ref.~\cite{WuChen2024}, we discuss the relationship between the double and triple decompositions of the vorticity field. On one hand, according to the normal-nilpotent decomposition (NND) of the VGT, the vorticity field can be doubly decomposed as $\bm{\omega}=2\bm{\psi}_{N}+\bm{s}_{N}$~\cite{LiuCQ2018,LiZhen2024}, where $\bm{\psi}_{N}$ and $\bm{s}_{N}$ represent the characteristic orbital rotation and spin used in the NND, respectively. $\bm{\psi}_{N}$ is parallel to the eigenvector $\bm{e}_{3}$ that corresponds to the unique real eigenvalue for $\Delta>0$ ($\Delta$ is the discriminant of the velocity gradient tensor). Unfortunately, the explicit expressions of $\bm{\psi}_{N}$ and $\bm{s}_{N}$ cannot be obtained under the framework of matrix algebra.
On the other hand, a triple decomposition of the vorticity is found as $\bm{\omega}=2\bm{\psi}+\bm{\omega}_{h}+(\bm{\omega}_{\rm{II}}-\bm{\psi})$. The role of the streaming vorticity $\bm{\omega}_{h}$ is turning the directions of both $\bm{R}\equiv2\bm{\psi}$ and $(\bm{\omega}_{\rm{II}}-\bm{\psi})$ to that of $\bm{e}_{3}$. The result provides a physical interpretation for the unitary and orthogonal transformations used in the NND.
} under which the total vorticity is the sum of the orbital rotation and the (generalized) spin. As an example, for the 2D circular motion with the velocity $\bm{u}=q(r)\bm{e}_{\theta}$, it is obvious that $\hat{\bm{t}}=\bm{e}_{\theta}$, $\hat{\bm{n}}=-\bm{e}_{r}$ and $\hat{\bm{b}}=\bm{e}_{z}$, where $\left\{\bm{e}_{r},\bm{e}_{\theta},\bm{e}_{z}\right\}$ are the base vectors corresponding to the cylindrical coordinate system $(r,\theta,z)$. Then, one can obtain $\bm{\psi}=r^{-1}q\bm{e}_{z}$ and $\bm{s}=(\partial_{r}q-r^{-1}q)\bm{e}_{z}$, which implies that $\omega=\partial_{r}q+r^{-1}q=r^{-1}\partial_{r}(rq)$.

It is claimed that the decomposition in Eqs.~\eqref{ORS} and~\eqref{RS} actually adds a new physical dimension to the vorticity dynamics in both the fluid interior and the boundary, which naturally results in a decomposition of the BVF as
\begin{subequations}
	\begin{eqnarray}\label{BVF_LPWu}
		\bm{\sigma}=2\bm{\sigma}_{\bm{\psi}}+\bm{\sigma}_{\bm{s}}=\bm{\sigma}_{\bm{R}}+\bm{\sigma}_{\bm{s}},~~\bm{\sigma}_{\bm{R}}\equiv2\bm{\sigma}_{\bm{\psi}},
	\end{eqnarray}
	\begin{eqnarray}\label{BVF_LPWu1}
		\bm{\sigma}_{\bm{\psi}}\equiv\nu\left[\partial_{n}\bm{\psi}\right]_{w},~\bm{\sigma}_{\bm{s}}\equiv\nu\left[\partial_{n}\bm{s}\right]_{w},
	\end{eqnarray}
\end{subequations}
where $\bm{\sigma}_{\bm{\psi}}$ and $\bm{\sigma}_{\bm{s}}$ are referred to as~\textit{the boundary orbital-rotation flux} and~\textit{the boundary spin flux}, respectively. 
Since the vorticity and its two vector components cannot penetrate into the solid side of the boundary,
their boundary fluxes become the physical measures of their boundary sources.
To the best of the authors' knowledge, a general vorticity-dynamical theory was never found in the existing literature that revealed the physical structures of the boundary fluxes (sources) of orbital rotation and spin, which becomes the core of the article.

The paper is organized as follows. In $\S$~\ref{NTE}, the near-wall Taylor-series expansion solutions of the velocity and its magnitude are obtained, which are used to evaluate the limits and the wall-normal derivatives in $\S$~\ref{LWD}. Then, the decompositions of the BVF and BEF are derived and interpreted in $\S$~\ref{DBVF} and $\S$~\ref{DBEF}, respectively. As an alternative description to the boundary vorticity dynamics, the decompositions of the Lyman vorticity and enstrophy fluxes are discussed in $\S$~\ref{DLVF} and $\S$~\ref{DLEF}. Conclusions and discussions are given in $\S$~\ref{Conclusions and discussions}.

\section{Near-wall Taylor-series expansion}\label{NTE}
The NS equations for incompressible viscous flow are written as
\begin{subequations}\label{NS}
\begin{eqnarray}\label{eq1}
\bm{\nabla}\bm{\cdot}\bm{u}=0,
\end{eqnarray}
\begin{eqnarray}\label{eq2}
	D_{t}\bm{u}=-\bm{\nabla}\hat{p}+\nu\nabla^{2}\bm{u},
\end{eqnarray}
\end{subequations}
where ${\rho}$ is the density, $p$ is the pressure and $D_{t}\bm{u}=D\bm{u}/Dt$ is the acceleration of a fluid particle. When the conservative body force (per unit mass) $\bm{f}=\bm{\nabla}\phi$ is considered, $p$ should be replaced by $p-\rho\phi$ with $\phi$ being the force 
potential. For convenience, we use $\hat{p}\equiv{p}/\rho$ in the following discussion.
\begin{figure}[t]
	\centering
	\includegraphics[width=0.9\columnwidth,trim={1.0cm 3.0cm 1.0cm 3.8cm},clip]{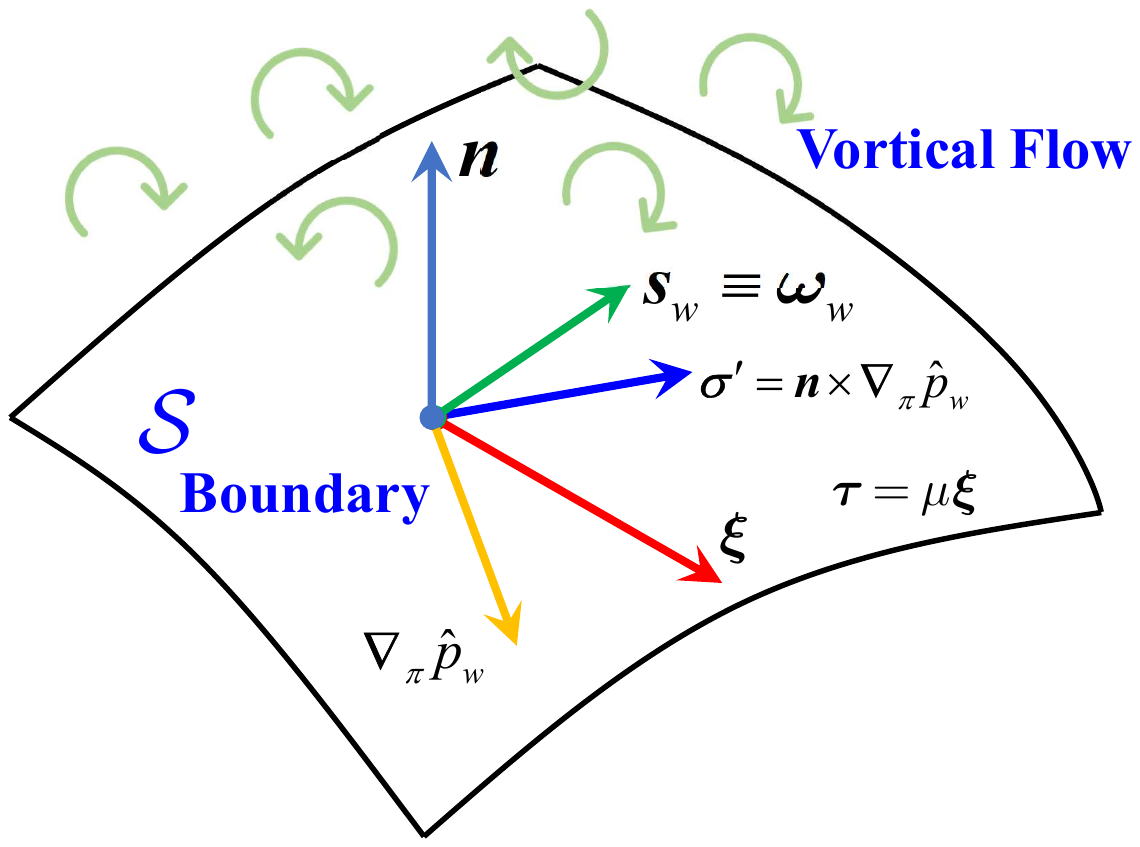}
	\caption{Sketch of the vortical flow interacting with a stationary solid boundary $\mathcal{S}$. Two on-wall orthogonal pairs, including $(\bm{\xi},\bm{s}_{w})$ and $(\bm{\nabla}_{\pi}\hat{p}_{w},\bm{\sigma}^{\prime})$, constituted by the elementary surface physical quantities are displayed with the arrows in different colors, where $\bm{s}_{w}$ is the spin on the wall (or the surface vorticity $\bm{\omega}_{w}$); $\bm{\xi}$ is proportional to the skin friction $\bm{\tau}$, being perpendicular to $\bm{s}_{w}$; $\bm{\nabla}_{\pi}\hat{p}_{w}$ is the surface tangential pressure gradient and $\bm{\sigma}^{\prime}$ is the Lyman vorticity flux (Eq.~\eqref{Lyman_flux_total}). $\bm{n}$ represents the unit normal vector of the boundary directed into the fluid.} 
	\label{Flow_Boundary}
\end{figure}

By applying the Taylor-series expansion to Eqs.~\eqref{eq1} and~\eqref{eq2} and using the no-slip boundary condition, it can be shown that the velocity in a small vicinity of the wall can be explicitly expressed by the fundamental surface physical quantities:
\begin{subequations}\label{Taylor_expansion}
\begin{eqnarray}\label{u_pi}
\bm{u}_{\pi}=\bm{\xi}y+\left(\frac{1}{2\nu}\bm{\nabla}_{\pi}\hat{p}_{w}+\frac{1}{2}K\bm{\xi}\right)y^2+\bm{\mathcal{O}}(y^3),
\end{eqnarray}
\begin{eqnarray}\label{u_n}
	{u}_{n}=-\frac{1}{2}\left(\bm{\nabla}_{\pi}\bm{\cdot}\bm{\xi}\right)y^2+\mathcal{O}(y^3).
\end{eqnarray}
\end{subequations}
where $\bm{u}_{\pi}$ and $u_{n}$ represent the tangential and wall-normal velocity components, respectively.
$y$ is the wall-normal coordinate measured from the wall ($y=0$).
$\bm{\nabla}_{\pi}\equiv\left[\bm{\nabla}-\bm{n}\partial_{n}\right]_{w}$ denotes the surface tangential gradient operator, and $\partial_{n}\equiv\partial_{y}$ is the wall-normal derivative ($\partial_{n}^{2}\equiv\partial_{n}\circ\partial_{n}$). $\bm{K}\equiv-\bm{\nabla}_{\pi}\bm{n}$ is the surface curvature tensor and $K\equiv tr(\bm{K})=-\bm{\nabla}_{\pi}\bm{\cdot}\bm{n}$ is twice of the mean curvature of the boundary surface.
On a stationary wall, as shown in Fig.~\ref{Flow_Boundary}, the orbital rotation vanishes ($\bm{\psi}_{w}=\bm{0}$) so that the vorticity is equal to the spin ($\bm{\omega}_{w}=\bm{s}_{w}$); $\bm{\xi}=\bm{s}_{w}\times\bm{n}$ is a vector perpendicular to the spin $\bm{s}_{w}$.
$\bm{\tau}=\mu\bm{\xi}$ is the skin friction ($\mu$ is the dynamic viscosity), which represents the 
wall shear stress exerted by the outer fluid.
From Eqs.~\eqref{u_pi} and~\eqref{u_n}, the velocity magnitude can be expanded as
\begin{eqnarray}\label{aa1}
q=\lVert\bm{\xi}\rVert y+\left(\frac{1}{2\nu}\partial_{\xi}\hat{p}_{w}+\frac{1}{2}K\lVert\bm{\xi}\rVert\right)y^2+\mathcal{O}(y^3).
\end{eqnarray}
Throughout the paper, $s_{\xi}$ and $s_{\omega}$ denote the arc length parameter along a $\bm{\xi}$-line and
a $\bm{s}_{w}$-line lying on the wall, respectively.
A special on-wall orthonormal frame $(\bm{e}_{\xi}, \bm{e}_{\omega}, \bm{n})$ is introduced where the base vectors are defined as $\bm{e}_{\xi}\equiv\bm{\xi}/\lVert\bm{\xi}\rVert$,
$\bm{e}_{\omega}\equiv\bm{s}_{w}/\lVert\bm{s}_{w}\rVert$ and $\bm{n}=\bm{e}_{\xi}\times\bm{e}_{\omega}$, respectively. For neatness, we use the following notations to denote the partial derivatives along the coordinate lines:
\begin{eqnarray}
\partial_{\xi}\equiv\frac{\partial}{\partial{s}_{\xi}}=\bm{e}_{\xi}\bm{\cdot}\bm{\nabla}_{\pi}~~\text{and}~~\partial_{\omega}\equiv\frac{\partial}{\partial{s}_{\omega}}=\bm{e}_{\omega}\bm{\cdot}\bm{\nabla}_{\pi}.
\end{eqnarray}
Under this reference frame, the surface curvature tensor $\bm{K}$ can be expanded as ($b_{\xi\omega}=b_{\omega\xi}$)
\begin{eqnarray}
\bm{K}=b_{\xi\xi}\bm{e}_{\xi}\bm{e}_{\xi}+b_{\omega\omega}\bm{e}_{\omega}\bm{e}_{\omega}+b_{\xi\omega}\bm{e}_{\xi}\bm{e}_{\omega}+b_{\omega\xi}\bm{e}_{\omega}\bm{e}_{\xi}.
\end{eqnarray}

\section{Limits and wall-normal derivatives}\label{LWD}
The on-wall values of $\hat{\bm{t}}$ and its wall-normal derivative $\left[\partial_{n}\hat{\bm{t}}\right]_{w}$ can be defined by their limits as $y\rightarrow{0}$. By using Eqs.~\eqref{Taylor_expansion} and~\eqref{aa1}, they can be evaluated as
\begin{subequations}
\begin{eqnarray}\label{m1}
	\hat{\bm{t}}_{w}\equiv\lim_{y\rightarrow0}\hat{\bm{t}}=\bm{e}_{\xi},
\end{eqnarray}
\begin{eqnarray}\label{m2}
	\left[\partial_{n}\hat{\bm{t}}\right]_{w}\equiv\lim_{y\rightarrow0}\partial_{n}\hat{\bm{t}}
=\left[\frac{1}{2\nu}\bm{e}_{\omega} \partial_{\omega}\hat{p}_{w}-\frac{1}{2}\left(\bm{\nabla}_{\pi}\bm{\cdot}\bm{\xi}\right)\bm{n}\right]\frac{1}{\lVert\bm{\xi}\rVert}.
\end{eqnarray}	
\end{subequations}

It is noted that the wall-normal pressure gradient at the wall is directly related to the skin friction divergence: $\left[\partial_{n}p\right]_{w}=-\bm{\nabla}_{\pi}\bm{\cdot}\bm{\tau}$ with $\bm{\tau}=\mu\bm{\xi}$.
Therefore, Eq.~\eqref{m2} indicates that the direction of the unit tangent vector of a streamline can be changed by the tangential surface pressure gradient along a vorticity line (or $\bm{s}_{w}$-line) and the wall-normal pressure gradient at the wall, which is also influenced by the distribution of the vorticity magnitude (or the enstrophy).
The derivation details of Eq.~\eqref{m2} are documented in~\ref{AppendixA}.
In addition, by comparing Eq.~\eqref{aa1} and the general form of the Taylor-series expansion, the first and second wall-normal derivatives of the velocity magnitude at the wall is directly obtained as
\begin{subequations}
\begin{eqnarray}\label{m3}
	\left[\partial_{n}q\right]_{w}=\lVert\bm{\xi}\rVert=\lVert\bm{s}_{w}\rVert=\lVert\bm{\omega}_{w}\rVert,
\end{eqnarray}
\begin{eqnarray}\label{m4}
	\left[\partial_{n}^{2}q\right]_{w}=\frac{1}{\nu}\partial_{\xi}\hat{p}_{w}+K\lVert\bm{s}_{w}\rVert.
\end{eqnarray}
\end{subequations}

\section{Decomposition of boundary vorticity flux}\label{DBVF}
The boundary fluxes $\bm{\sigma}_{\rm{I}}\equiv\nu[\partial_{n}\bm{\omega}_{\rm{I}}]_{w}$ and $\bm{\sigma}_{\rm{II}}\equiv\nu[\partial_{n}\bm{\omega}_{\rm{II}}]_{w}$ are derived as (\ref{new81} and~\ref{new82})
\begin{subequations}
	\begin{eqnarray}\label{BFI}
		\bm{\sigma}_{\rm{I}}=\nu{\partial_{\omega}\lVert\bm{s}_{w}\rVert}\bm{n}-\frac{1}{2}\bm{e}_{\xi}{\partial_{\omega}\hat{p}_{w}}+\nu\bm{K}\bm{\cdot}\bm{s}_{w}-\nu K\bm{s}_{w}-\nu\left(\bm{\nabla}_{\pi}\bm{\cdot}\bm{s}_{w}\right)\bm{n}.
	\end{eqnarray}
	\begin{eqnarray}\label{BFII}
		\bm{\sigma}_{\rm{II}}=-\nu{\partial_{\omega}\lVert\bm{s}_{w}\rVert}\bm{n}+\frac{1}{2}\bm{e}_{\xi}\partial_{\omega}\hat{p}_{w}+\bm{n}\times\bm{\nabla}_{\pi}\hat{p}_{w}+\nu{K}\bm{s}_{w},
	\end{eqnarray}
\end{subequations}
respectively. The sum of $\bm{\sigma}_{\rm{I}}$ and $\bm{\sigma}_{\rm{II}}$ yields exactly the total BVF:
\begin{eqnarray}\label{total_BVF}
\color{red}\bm{\sigma}=\bm{n}\times\bm{\nabla}_{\pi}\hat{p}_{w}+\nu\bm{K}\bm{\cdot}\bm{s}_{w}-\nu\left(\bm{\nabla}_{\pi}\bm{\cdot}\bm{s}_{w}\right)\bm{n}.
\end{eqnarray}
By using the method of differential geometry, compact expressions of $\bm{\sigma}_{\rm{I}}$, $\bm{\sigma}_{\rm{II}}$ and $\bm{\sigma}$ are obtained in~\ref{BVFsCR} where we find that the effective contributions of $\bm{\sigma}_{\rm{I}}$ and $\bm{\sigma}_{\rm{II}}$ to $\bm{\sigma}$ only come from the terms related to the wedge product and the Hodge star operator. In the right hand side of Eq.~\eqref{total_BVF}, the first term represents the vorticity creation by the surface pressure gradient, which is the dominant mechanism at the high-Reynolds-number limit; the second term is interpreted as the spin-curvature coupling; the third term is just the wall-normal BVF, being associated with the surface spin divergence.

The orbital rotation $\bm{\psi}$ in Eq.~\eqref{orbital_rotation} is based on the quantities defined on a streamline, which can be equivalently written in terms of $\hat{\bm{t}}$ and the VGT $\bm{A}\equiv\bm{\nabla}\bm{u}$ (\ref{AppendixB})
\begin{eqnarray}\label{BOF1}
	\bm{\psi}=\hat{\bm{t}}\times\left(\hat{\bm{t}}\bm{\cdot}\bm{A}\right)=\frac{1}{q^2}{\bm{u}}\times\left({\bm{u}}\bm{\cdot}\bm{A}\right).
\end{eqnarray}
Equation~\eqref{BOF1} provides a $\hat{\bm{t}}$-dependent intrinsic representation for the orbital rotation, which holds for any complex flows. To the best of the authors’ knowledge, it has not previously been found in the existing literature, by which the (generalized) spin is derived as
\begin{eqnarray}
\bm{s}=\left(\bm{\omega}\bm{\cdot}\hat{\bm{t}}\right)\hat{\bm{t}}-2\hat{\bm{t}}\times\left(\hat{\bm{t}}\bm{\cdot}\bm{D}\right)=\frac{1}{q^2}\left[\left(\bm{\omega}\bm{\cdot}\bm{u}\right)\bm{u}-2\bm{u}\times\left(\bm{u}\bm{\cdot}\bm{D}\right)\right],
\end{eqnarray}
where $\bm{D}=(\bm{A}+\bm{A}^{T})/2$ is the strain-rate tensor. The term $\hat{\bm{t}}\times\left(\hat{\bm{t}}\bm{\cdot}\bm{D}\right)$ represents the angular velocity describing the specific rotation rate of a line element $\delta\bm{r}=\hat{\bm{t}}{\delta}r$~\cite{WuChen2024}.

From Eq.~\eqref{BOF1}, it is proved that the boundary orbital-rotation flux $\bm{\sigma}_{\bm{\psi}}$ is expressed as (\ref{boundary_OR_flux_derivation})
\begin{eqnarray}\label{sigma_psi}
	\bm{\sigma}_{\bm\psi}=\nu\bm{e}_{\xi}\times\left(\bm{e}_{\xi}\bm{\cdot}\bm{\nabla}_{\pi}\bm{\xi}\right)
	=\nu\kappa_{g,\xi}\lVert\bm{s}_{w}\rVert\bm{n}-\nu{b}_{\xi\xi}\bm{s}_{w},
\end{eqnarray}
where $\kappa_{g,\xi}$ is the geodesic curvature of the $\bm{\xi}$-line (or the skin friction line). However, we notice that it is usually difficult to calculate $\kappa_{g,\xi}$ from the simulation or experiment datasets. To overcome this difficulty, by using Eqs.~\eqref{qqq5} and~\eqref{qqq7}, Eq.~\eqref{sigma_psi} can be transformed into
\begin{eqnarray}\label{sigma_psi2}
	\color{red}\bm{\sigma}_{\bm\psi}
	=\nu\left(\partial_{\omega}\lVert\bm{s}_{w}\rVert-\bm{\nabla}_{\pi}\bm{\cdot}\bm{s}_{w}\right)\bm{n}-\nu{b}_{\xi\xi}\bm{s}_{w}.
\end{eqnarray}
Each term in Eq.~\eqref{sigma_psi2} can be expanded under a given curvilinear coordinate system in practical computation.
Interestingly, $\bm{\sigma}_{\bm\psi}$ has zero component in the $\bm{e}_{\xi}$-direction. In other words, the orbital rotation created on the wall diffuses inside the plane spanned by $\bm{n}$ and $\bm{s}_{w}$.
The wall-normal component ($\bm{\sigma}_{\bm\psi}\bm{\cdot}\bm{n}=\nu\kappa_{g,\xi}\lVert\bm{s}_{w}\rVert$) is determined by the coupling between the geodesic curvature and the vorticity magnitude (multiplied by $\nu$), which is a part of the wall-normal BVF ($\bm{\sigma}\bm{\cdot}\bm{n}=-\nu\bm{\nabla}_{\pi}\bm{\cdot}\bm{s}_{w}=-\nu\partial_{\omega}\lVert\bm{s}_{w}\rVert+\nu\kappa_{g,\xi}\lVert\bm{s}_{w}\rVert$). The tangential component ($\bm{\sigma}_{\bm\psi}\bm{\cdot}\bm{e}_{\omega}=-\nu{b}_{\xi\xi}\lVert\bm{s}_{w}\rVert$) is parallel to the spin, which represents a pure geometrical effect on a curved surface. Therefore, $\bm{\sigma}_{\bm\psi}$ must direct towards the wall-normal direction for a flat boundary while in the presence of a curved surface, the curvature effect can make the orbital rotation deviate from the wall-normal direction.

The difference between Eqs.~\eqref{total_BVF} and~\eqref{sigma_psi2} ($\times$2) yields the boundary spin flux $\bm{\sigma}_{\bm{s}}$:
\begin{eqnarray}\label{sigma_spin}
\bm{\sigma}_{\bm{s}}&=&\bm{n}\times\bm{\nabla}_{\pi}\hat{p}_{w}
+\nu{b}_{\xi\omega}\bm{\xi}+\nu(2b_{\xi\xi}+b_{\omega\omega})\bm{s}_{w}\nonumber\\
& &-\nu(\partial_{\omega}\lVert\bm{s}_{w}\rVert+\kappa_{g,\xi}\lVert\bm{s}_{w}\rVert)\bm{n}.
\end{eqnarray}
In the right hand side of Eq.~\eqref{sigma_spin}, the first term indicates that $\bm{\sigma}_{\bm s}$ can be created by the surface pressure gradient which is not responsible for the generation of $\bm{\sigma}_{\bm\psi}$ in Eq.~\eqref{sigma_psi2}. In other words, the boundary coupling between the transverse and longitudinal fields is only embodied in the creation of spin on a boundary.
The second term represents the off-diagonal element of the surface curvature tensor ($b_{\xi\omega}$) coupled with the transverse field ($\bm{\xi}$). The third term represents the coupling between the surface geometry ($K+b_{\xi\xi}=2b_{\xi\xi}+b_{\omega\omega}$) and the spin ($\bm{s}_{w}$). The fourth term is related to the wall-normal BVF while the sign is changed for the term $\nu\kappa_{g,\xi}\lVert\bm{s}_{w}\rVert$.
Under the on-wall orthonormal frame $(\bm{e}_{\xi},\bm{e}_{\omega},\bm{n})$, the three components of $\bm{\sigma}_{\bm{s}}$ are explicitly written as
\begin{subequations}
\begin{equation}
	\bm{\sigma}_{\bm{s}}\bm{\cdot}\bm{e}_{\xi}=-\partial_{\omega}\hat{p}_{w}+\nu{b}_{\xi\omega}\lVert\bm{s}_{w}\rVert,
\end{equation}
\begin{equation}
	\bm{\sigma}_{\bm{s}}\bm{\cdot}\bm{e}_{\omega}=\partial_{\xi}\hat{p}_{w}+\nu(2b_{\xi\xi}+b_{\omega\omega})\lVert\bm{s}_{w}\rVert,
\end{equation}
\begin{equation}
	\bm{\sigma}_{\bm{s}}\bm{\cdot}\bm{n}=-\nu(\partial_{\omega}\lVert\bm{s}_{w}\rVert+\kappa_{g,\xi}\lVert\bm{s}_{w}\rVert).
\end{equation}
\end{subequations}

By using Eqs.~\eqref{spin},~\eqref{BFII} and~\eqref{sigma_spin}, the boundary streaming vorticity flux $\bm{\sigma}_{h}\equiv\nu[\partial_{n}\bm{\omega}_{h}]_{w}$ is obtained as
\begin{eqnarray}\label{sigma_h}
\bm{\sigma}_{h}=-\frac{1}{2}\bm{e}_{\bm{\xi}}\partial_{\omega}\hat{p}_{w}+\nu{b}_{\xi\omega}\bm{\xi}.
\end{eqnarray}
The first term in the right hand side of Eq.~\eqref{sigma_h} suggests that the streaming vorticity can be created by the surface pressure gradient along the vorticity line (namely, $\bm{e}_{\bm\omega}$-direction) which then diffuses along the skin friction line (namely, $\bm{e}_{\bm\xi}$-direction).
The other boundary source of the streaming vorticity is $\nu{b}_{\xi\omega}\bm{\xi}$, representing the coupling between the spin and the surface geometry, where $b_{\xi\omega}$ is shown to be the geodesic torsion of the skin friction line. Obviously, $\bm{\sigma}_{h}$ vanishes in 2D cases, thereby being identified as a pure 3D physical mechanism.

\section{Decomposition of boundary enstrophy flux}\label{DBEF}
In this section, we extend the discussion of the boundary vorticity dynamics to the boundary enstrophy dynamics.
Due to Eq.~\eqref{ORS}, the total enstrophy $\Omega$ can be decomposed into three terms, namely,
\begin{subequations}
	\begin{eqnarray}
		\Omega\equiv\frac{1}{2}\omega^{2}=\Omega_{R}+\Omega_{Rs}+\Omega_{s},
	\end{eqnarray}
	\begin{eqnarray}
		\Omega_{R}\equiv\frac{1}{2}R^2=2\psi^{2},~\Omega_{R s}\equiv\bm{R}\bm{\cdot}\bm{s},
		~\Omega_{s}\equiv\frac{1}{2}s^2,
	\end{eqnarray}
\end{subequations}
where $\Omega_{R}$ and $\Omega_{s}$ are the enstrophy caused by the orbital rotation $\bm{\psi}$ and the spin $\bm{s}$, respectively; $\Omega_{R{s}}$ represents the coupling effect between $\bm{\psi}$ and $\bm{s}$. Correspondingly, the BEF $F_{\Omega}$ is split as the sum of three contributions:
	\begin{eqnarray}
F_{\Omega}\equiv\nu\left[\partial_{n}\Omega\right]_{w}=\bm{\omega}_{w}\bm{\cdot}\bm{\sigma}=F_{R}+F_{Rs}+F_{s},
	\end{eqnarray}
where $F_{R}\equiv\nu\left[\partial_{n}\Omega_{R}\right]_{w}$, $F_{Rs}\equiv\nu\left[\partial_{n}\Omega_{Rs}\right]_{w}$ and $F_{s}\equiv\nu\left[\partial_{n}\Omega_{s}\right]_{w}$ are referred to as the boundary $\Omega_{R}$ flux, the boundary $\Omega_{Rs}$ flux and the boundary $\Omega_{s}$ flux, respectively.
On a stationary wall, since the orbital rotation disappears ($\bm{\psi}_{w}=\bm{0}$), no boundary $\Omega_{R}$ flux is created ($F_{R}=0$) and the boundary $\Omega_{Rs}$ flux is simplified as $F_{Rs}=\bm{s}_{w}\bm{\cdot}\bm{\sigma}_{\bm{R}}$. The boundary $\Omega_{s}$ flux is related to the boundary spin flux through $F_{s}=\bm{s}_{w}\bm{\cdot}\bm{\sigma}_{\bm s}$.
	
From Eqs.~\eqref{sigma_psi2} and~\eqref{sigma_spin}, it is shown that $F_{Rs}$ and $F_{s}$ are 
\begin{subequations}
	\begin{eqnarray}\label{XX1}
	\color{red}F_{Rs}=-2\nu\bm{\xi}\bm{\cdot}\bm{K}\bm{\cdot}\bm{\xi},
	\end{eqnarray}
	\begin{eqnarray}\label{XX2}
	\color{red}	F_{s}=\bm{\xi}\bm{\cdot}\bm{\nabla}_{\pi}\hat{p}_{w}+\nu\bm{s}_{w}\bm{\cdot}\bm{K}\bm{\cdot}\bm{s}_{w}+2\nu\bm{\xi}\bm{\cdot}\bm{K}\bm{\cdot}\bm{\xi}.
	\end{eqnarray}
\end{subequations}
The sum of Eqs.~\eqref{XX1} and~\eqref{XX2} gives the BEF $F_{\Omega}$:
\begin{eqnarray}\label{XX3}
\color{red}	F_{\Omega}=\bm{\xi}\bm{\cdot}\bm{\nabla}_{\pi}\hat{p}_{w}+\nu\bm{s}_{w}\bm{\cdot}\bm{K}\bm{\cdot}\bm{s}_{w}.
\end{eqnarray}
Equation~\eqref{XX1} show that $F_{Rs}$ can only be created by the quadratic form $\bm{\xi}\bm{\cdot}\bm{K}\bm{\cdot}\bm{\xi}=2b_{\xi\xi}\Omega_{w}$ representing the spin-curvature coupling effect. Interestingly, the same quadratic term also contributes to $F_{s}$ in Eq.~\eqref{XX2} although with the opposite sign.
Therefore, the cancellation between these two terms makes $\bm{\xi}\bm{\cdot}\bm{K}\bm{\cdot}\bm{\xi}$ ineffective in generating the total BEF $F_{\Omega}$.
The other quadratic term, $\bm{s}_{w}\bm{\cdot}\bm{K}\bm{\cdot}\bm{s}_{w}=2b_{\omega\omega}\Omega_{w}$, only results in creating the boundary spin flux. The boundary coupling of the longitudinal and transverse processes is described by the dot product between $\bm{\xi}$ and the surface pressure gradient $\bm{\nabla}_{\pi}\hat{p}_{w}$, which becomes another physical mechanism in generating $F_{s}$. Owing to the intrinsic constraints of geometry, these two quadratic terms can be related to the geometrical properties of the curves lying on the surface.~\footnote{
	The integral lines of $\bm{\xi}$ and $\bm{s}_{w}$ form a dense network covering the surface. For a fixed point on the surface, there exist a $\bm{\xi}$-line and a $\bm{s}_{w}$-line 
	being tangent to $\bm{e}_{\bm\xi}$ and $\bm{e}_{\bm{\omega}}$ defined at that point, respectively. By applying the method of the differential geometry, it is direct to show that 
	${\kappa}_{\xi}(\bm{n}_{\xi}\bm{\cdot}\bm{n})=b_{\xi\xi}$ and 
	${\kappa}_{\omega}(\bm{n}_{\omega}\bm{\cdot}\bm{n})=b_{\omega\omega}$,
	where $\bm{n}_{\xi}$ ($\bm{n}_{\omega}$) and $\kappa_{\xi}$ ($\kappa_{\omega}$) are the principal unit normal vector and the curvature of the $\bm{\xi}$-line ($\bm{s} _{w}$-line), respectively. Equivalently, the two quadratic forms can be written as $\bm{\xi}\bm{\cdot}\bm{K}\bm{\cdot}\bm{\xi}=2\kappa_{\xi}(\bm{n}_{\xi}\bm{\cdot}\bm{n})\Omega_{w}$ and $\bm{s}_{w}\bm{\cdot}\bm{K}\bm{\cdot}\bm{s}_{w}=2\kappa_{\omega}(\bm{n}_{\omega}\bm{\cdot}\bm{n})\Omega_{w}$.} Interestingly, although the streaming vorticity $\bm{\omega}_{h}$ is indispensible in analyzing the decomposition of the vorticity in the fluid interior, it does not affect the boundary enstrophy dynamics: both $F_{s}$ and $F_{\Omega}$ are not contributed by $\bm{\sigma}_{h}$ owing to the orthogonality of $\bm{s}_{w}$ and $\bm{\sigma}_{h}$.

As the second invariant of the velocity gradient tensor, $Q$ represents 
the source terms of the Phillips equation of the density fluctuation for the weakly compressible flow and the Poisson equation for incompressible flow. Therefore, revealing the boundary creation mechanism of $Q$ is of crucial importance to the control of flow noise on the boundary.
Recently, Chen et al.~\cite{ChenWu2024} have shown that the boundary $Q$ flux is expressed as
\begin{eqnarray}\label{FQ}
	F_{Q}\equiv\nu\left[\partial_{n}Q\right]_{w}=-\nu\bm{\xi}\bm{\cdot}\bm{K}\bm{\cdot}\bm{\xi}.
\end{eqnarray}
Equation~\eqref{FQ} implies that for incompressible viscous flow interacting with a stationary wall, the unique boundary source of $Q$ is the boundary coupling between the longitudinal field ($\bm{\xi}$) and the surface geometry ($\bm{K}$).
Comparing Eqs.~\eqref{XX1} and~\eqref{FQ} gives $F_{Rs}=2F_{Q}$
which implies that the created boundary $Q$ flux induces the interaction between the orbital rotation and the spin in a vicinity of the wall. This interaction continues and gives rise to the change of the curvature of a shear layer, thereby initiating the rolling-up process to form an axial vortex above the wall.

\section{Decomposition of Lyman vorticity flux}\label{DLVF}
The Lyman vorticity flux $\bm{\sigma}^{\prime}\equiv-\nu\left[\bm{n}\times\left(\bm{\nabla}\times\bm{\omega}\right)\right]_{w}$ is an alternative definition of the boundary vorticity flux, which represents the tangential component of the viscous force term in Eq.~\eqref{eq2}~\cite{Lyman1990}. 
The integral equivalence of $\bm{\sigma}$ and $\bm{\sigma}^{\prime}$ over a closed surface $\mathcal{S}$ can be easily validated through the generalized Stokes theorem, namely,
\begin{eqnarray}
	\oint_{\mathcal{S}}\bm{\sigma}d\mathcal{S}=\oint_{\mathcal{S}}\bm{\sigma}^{\prime}d\mathcal{S}
	+\nu\oint_{\mathcal{S}}\left(\bm{n}\times\bm{\nabla}\right)\times\bm{\omega}d\mathcal{S}
	=\oint_{\mathcal{S}}\bm{\sigma}^{\prime}d\mathcal{S},
\end{eqnarray}
although they should be differentiated in describing the local vorticity creation rate on a boundary.

By Eq.~\eqref{OI_II}, the Lyman vorticity flux $\bm{\sigma}^{\prime}$ can be decomposed as
\begin{subequations}\label{Lyman_I_II}
	\begin{eqnarray}\label{Lyman_eq1}
		\bm{\sigma}^{\prime}=\bm{\sigma}^{\prime}_{\rm{I}}+\bm{\sigma}^{\prime}_{\rm{II}},
	\end{eqnarray}
where the Lyman $\bm{\omega}_{\rm{I}}$ flux and the Lyman $\bm{\omega}_{\rm{II}}$ flux are respectively defined by
	\begin{eqnarray}\label{Lyman_eq2}
		\bm{\sigma}^{\prime}_{\rm{I}}\equiv-\nu\left[\bm{n}\times\left(\bm{\nabla}\times\bm{\omega}_{\rm{I}}\right)\right]_{w}~~\text{and}~~
		~~\bm{\sigma}^{\prime}_{\rm{II}}\equiv-\nu\left[\bm{n}\times\left(\bm{\nabla}\times\bm{\omega}_{\rm{II}}\right)\right]_{w}.
	\end{eqnarray}
\end{subequations}
From Eq.~\eqref{omega12}, the full expressions of $\bm{\sigma}^{\prime}_{\rm{I}}$ and $\bm{\sigma}^{\prime}_{\rm{II}}$ are respectively derived as
~\footnote{Alternatively, Eqs.~\eqref{uu1} and~\eqref{uu2} can be written as
	\begin{eqnarray*}\label{uu1a}
		\bm{\sigma}_{\rm{I}}^{\prime}=\nu{b}_{\xi\omega}\bm{\xi}-\nu{b}_{\xi\xi}\bm{s}_{w}-\frac{1}{2}\bm{e}_{\xi}\partial_{\omega}\hat{p}_{w},
	\end{eqnarray*}
	\begin{eqnarray*}\label{uu2a}
		\bm{\sigma}_{\rm{II}}^{\prime}=-\nu{b}_{\xi\omega}\bm{\xi}+\nu{b}_{\xi\xi}\bm{s}_{w}+\frac{1}{2}\bm{e}_{\xi}\partial_{\omega}\hat{p}_{w}+\bm{n}\times\bm{\nabla}_{\pi}\hat{p}_{w}.
	\end{eqnarray*}}
\begin{subequations}\label{uu12}
\begin{eqnarray}\label{uu1}
\bm{\sigma}_{\rm{I}}^{\prime}=\nu\bm{K}\bm{\cdot}\bm{s}_{w}-\nu{K}\bm{s}_{w}-\frac{1}{2}\bm{e}_{\xi}\partial_{\omega}\hat{p}_{w},
\end{eqnarray}
\begin{eqnarray}\label{uu2}
\bm{\sigma}_{\rm{II}}^{\prime}=-\nu\bm{K}\bm{\cdot}\bm{s}_{w}+\nu{K}\bm{s}_{w}+\frac{1}{2}\bm{e}_{\xi}\partial_{\omega}\hat{p}_{w}+\bm{n}\times\bm{\nabla}_{\pi}\hat{p}_{w}.
\end{eqnarray}
\end{subequations}
The sum of Eqs.~\eqref{uu1} and~\eqref{uu2} yields the Lyman vorticity flux:
\begin{eqnarray}\label{Lyman_flux_total}
	\color{red}\bm{\sigma}^{\prime}=\bm{n}\times\bm{\nabla}_{\pi}\hat{p}_{w},
\end{eqnarray}
which is completely determined by the tangential surface pressure gradient on a stationary wall.

By using Eq.~\eqref{ORS}, the Lyman vorticity flux $\bm{\sigma}^{\prime}\equiv-\nu\left[\bm{n}\times\left(\bm{\nabla}\times\bm{\omega}\right)\right]_{w}$ can be decomposed as 
\begin{subequations}
	\begin{eqnarray}
\bm{\sigma}^{\prime}=2\bm{\sigma}^{\prime}_{\bm{\psi}}+\bm{\sigma}^{\prime}_{\bm{s}}=\bm{\sigma}^{\prime}_{\bm{R}}+\bm{\sigma}^{\prime}_{\bm{s}},~~\bm{\sigma}^{\prime}_{\bm{R}}\equiv2\bm{\sigma}^{\prime}_{\bm{\psi}},
	\end{eqnarray}
	\begin{eqnarray}
		\bm{\sigma}^{\prime}_{\bm{\psi}}\equiv-\nu\left[\bm{n}\times\left(\bm{\nabla}\times\bm{\psi}\right)\right]_{w},
		~~\bm{\sigma}^{\prime}_{\bm{s}}\equiv-\nu\left[\bm{n}\times\left(\bm{\nabla}\times\bm{s}\right)\right]_{w},
	\end{eqnarray}
\end{subequations}
where $\bm{\sigma}^{\prime}_{\bm{\psi}}$ and $\bm{\sigma}^{\prime}_{\bm{s}}$ can be referred to as \textit{the Lyman orbital-rotation flux} and \textit{the Lyman spin flux}, respectively.

From Eq.~\eqref{RS}, the explicit expressions of $\bm{\sigma}^{\prime}_{\bm{\psi}}$ and $\bm{\sigma}^{\prime}_{\bm{s}}$ are respectively derived as
\begin{subequations}\label{Lyman_psi_sigma_flux}
	\begin{eqnarray}\label{Lyman_psi}
\color{red}\bm{\sigma}^{\prime}_{\bm{\psi}}=-\nu{b}_{\xi\xi}\bm{s}_{w}~~\text{or}~~\color{red}\bm{\sigma}^{\prime}_{\bm{R}}=-2\nu{b}_{\xi\xi}\bm{s}_{w},
	\end{eqnarray}
	\begin{eqnarray}\label{Lyman_s}
\color{red}\bm{\sigma}^{\prime}_{\bm{s}}=\bm{n}\times\bm{\nabla}_{\pi}\hat{p}_{w}+2\nu{b}_{\xi\xi}\bm{s}_{w}.
	\end{eqnarray}
\end{subequations}
We remark that $\bm{\sigma}^{\prime}_{\bm{\psi}}$ in Eq.~\eqref{Lyman_psi} actually represents the tangential component of the boundary orbital-rotation flux $\bm{\sigma}_{\bm{\psi}}$, which diffuses the created vorticity along the vorticity line. In contrast, $\bm{\sigma}^{\prime}_{\bm{s}}$ is not solely determined by the tangential component of the boundary spin flux $\bm{\sigma}_{\bm{s}}$, whereas an additional term $-\nu\bm{K}\bm{\cdot}\bm{s}_{w}$ should be added. The final expression of $\bm{\sigma}^{\prime}_{\bm{s}}$ in Eq.~\eqref{Lyman_s} indicates that the Lyman spin flux is generated by the tangential surface pressure gradient ($\bm{n}\times\bm{\nabla}_{\pi}\hat{p}_{w}$) and the coupling between the $\bm{e}_{\xi}$-$\bm{e}_{\xi}$ component of the surface curvature tensor (${b_{\xi\xi}}$) and the spin ($\bm{s}_{w}$).
Moreover, we find that the Lyman streaming vorticity flux $\bm{\sigma}^{\prime}_{h}\equiv-\nu\left[\bm{n}\times\left(\bm{\nabla}\times\bm{\omega}_{h}\right)\right]_{w}$ is exactly equal to the boundary streaming vorticity flux $\bm{\sigma}_{h}$ in Eq.~\eqref{sigma_h}. The sum of $\bm{\sigma}^{\prime}_{h}$ and $\bm{\sigma}^{\prime}_{\rm{II}}$ recovers $\bm{\sigma}^{\prime}_{\bm{s}}$ in Eq.~\eqref{Lyman_s}.

\section{Decomposition of Lyman enstrophy flux}\label{DLEF}
By using Eq.~\eqref{ORS}, the Lyman enstrophy flux $F_{\Omega}^{\prime}$ is decomposed as
\begin{eqnarray}\label{c1}
F^{\prime}_{\Omega}\equiv\bm{\omega}_{w}\bm{\cdot}\bm{\sigma}^{\prime}=F^{\prime}_{R}+F^{\prime}_{Rs}+F^{\prime}_{s},
\end{eqnarray}
where $F^{\prime}_{R}\equiv\bm{R}_{w}\bm{\cdot}\bm{\sigma}_{\bm{R}}^{\prime}$, $F^{\prime}_{Rs}\equiv\bm{s}_{w}\bm{\cdot}\bm{\sigma}_{\bm{R}}^{\prime}+\bm{R}_{w}\bm{\cdot}\bm{\sigma}_{\bm{s}}^{\prime}$ and $F_{s}^{\prime}\equiv\bm{s}_{w}\bm{\cdot}\bm{\sigma}^{\prime}_{\bm{s}}$ are referred to as the Lyman $\Omega_{R}$ flux, the Lyman $\Omega_{R s}$ flux and the Lyman $\Omega_{s}$ flux, respectively. Obviously, the Lyman $\Omega_{R}$ flux is $F_{R}^{\prime}=0$. 
Being similar to $\bm{\sigma}_{h}$, $\bm{\sigma}_{h}^{\prime}$ makes no contribution to both $F^{\prime}_{s}$ and $F^{\prime}_{\Omega}$.
By using Eqs.~\eqref{Lyman_psi} and~\eqref{Lyman_s}, the Lyman $\Omega_{Rs}$ flux and the Lyman $\Omega_{s}$ flux are respectively derived as
\begin{subequations}\label{c2}
\begin{eqnarray}\label{c23}
\color{red}F^{\prime}_{R s}=-2\nu\bm{\xi}\bm{\cdot}\bm{K}\bm{\cdot}\bm{\xi},
\end{eqnarray}
\begin{eqnarray}\label{c22}
\color{red}F_{s}^{\prime}=\bm{\xi}\bm{\cdot}\bm{\nabla}_{\pi}\hat{p}_{w}+2\nu\bm{\xi}\bm{\cdot}\bm{K}\bm{\cdot}\bm{\xi}.
\end{eqnarray}
\end{subequations}
It is claimed that $F^{\prime}_{Rs}$ in Eq.~\eqref{c23} is exactly the same as $F_{Rs}$ in Eq.~\eqref{XX1}. However, compared to $F_{s}$ in Eq.~\eqref{XX2}, the quadratic form $\bm{s}_{w}\bm{\cdot}\bm{K}\bm{\cdot}\bm{s}_{w}$ disappears in Eq.~\eqref{c22}.
Substituting Eq.~\eqref{c2} into Eq.~\eqref{c1} yields the Lyman enstrophy flux $F_{\Omega}^{\prime}$:
\begin{eqnarray}
\color{red}F^{\prime}_{\Omega}=\bm{\xi}\bm{\cdot}\bm{\nabla}_{\pi}\hat{p}_{w},
\end{eqnarray}
where is solely attributed to the coupling between longitudinal and transverse processes on a solid boundary.

\section{Conclusions and discussions}\label{Conclusions and discussions}
The present study is a new development to the fundamental theory of the \textit{boundary vorticity dynamics} which reveals the physical origins of the orbital rotation ($\bm{\psi}$ or $\bm{R}$) and the spin ($\bm{s}$) for incompressible viscous flow interacting with a stationary wall. The main findings and contributions are summarized as follows.

1. Starting from the intrinsic decomposition of the vorticity, the boundary vorticity flux (BVF) $\bm{\sigma}$, which describes the local creation rate of vorticity on a solid boundary,
is decomposed as the sum of twice of the boundary orbital-rotation flux ($\bm{\sigma}_{\bm\psi}$) and the boundary spin flux ($\bm{\sigma}_{\bm{s}}$). 
For the first time, general formulae of $\bm{\sigma}_{\bm\psi}$ and $\bm{\sigma}_{\bm{s}}$ are derived based on which all the boundary sources of the orbital-rotation ($\bm{\psi}$) and the spin ($\bm{s}$) are explicitly unraveled. The full expression of the boundary streaming vorticity flux ($\bm{\sigma}_{h}$) is also obtained, where the streaming vorticity ($\bm{\omega}_{h}$) is treated as a part of the (generalized) spin ($\bm{s}$).

2. We find that $\bm{\sigma}_{\bm\psi}$ is always perpendicular to the skin friction (or $\bm{\xi}$), which must diffuse in the normal plane of a skin friction line. For a flat boundary, $\bm{\sigma}_{\bm\psi}$ has only the wall-normal component as a part of the total BVF ($\bm{\sigma}$), being interpreted as the interaction between the geodesic curvature of a skin friction line and the spin magnitude. The direction of $\bm{\sigma}_{\bm\psi}$ could be changed by the curvature effect in the presence of a curved boundary. Interestingly, the boundary coupling between the transverse and longitudinal processes is not responsible for producing $\bm{\sigma}_{\bm\psi}$, which however, is only embodied in the generation of $\bm{\sigma}_{\bm{s}}$.

3. We have extended the decomposition of the BVF to the boundary enstrophy dynamics where general expressions of the boundary $\Omega_{Rs}$ flux ($F_{Rs}$) and the boundary $\Omega_{s}$ flux ($F_{s}$) are derived, respectively. The unique mechanism by which $F_{R{s}}$ can be created on a boundary is the quadratic form $-2\nu\bm{\xi}\bm{\cdot}\bm{K}\bm{\cdot}\bm{\xi}$, describing the interaction between the transverse field and the surface geometry.
Interestingly, it holds that $F_{Rs}=2F_{Q}$, where $F_{Q}$ is the boundary $Q$ flux ($Q$ represents the second principal invariant of the velocity gradient tensor). This fact implies that $F_{Q}$ could affect the $\bm{R}$-$\bm{s}$ interaction in a small vicinity of the wall, thereby initiating the rolling-up process of a shearing layer to form an axial vortex. $F_{s}$ can be generated by the boundary coupling of the longitudinal and transverse fields through the dot product ($\bm{\xi}\bm{\cdot}\bm{\nabla}_{\pi}\hat{p}_{w}$), the quadratic form ($2\nu\bm{\xi}\bm{\cdot}\bm{K}\bm{\cdot}\bm{\xi}$), and the other quadratic form ($\nu\bm{s}_{w}\bm{\cdot}\bm{K}\bm{\cdot}\bm{s}_{w}$). All the effective generation mechanisms to the total BEF are contributed by $F_{s}$.

4. While the streaming vorticity ($\bm{\omega}_{h}$) is of crucial importance in understanding the intrinsic splitting of vorticity, the boundary streaming flux $\bm{\sigma}_{h}$ does not produce any effective contribution to both the boundary $\Omega_{Rs}$ flux ($F_{Rs}$) and the boundary $\Omega_{s}$ flux ($F_{s}$).

5. Lyman vorticity flux ($\bm{\sigma}^{\prime}$) provides an alternative physical description to the boundary vorticity dynamics, which is distinguished from the BVF ($\bm{\sigma}$) in describing the local vorticity creation rate on a boundary. Similar to the decomposition of $\bm{\sigma}$, $\bm{\sigma}^{\prime}$ is split as the sum of twice of the Lyman orbital-rotation flux ($\bm{\sigma}_{\bm\psi}^{\prime}$) and the Lyman spin flux ($\bm{\sigma}_{\bm{s}}^{\prime}$). We find that $\bm{\sigma}_{\bm\psi}^{\prime}$ appears only on a curved boundary, which diffuses the vorticity along the surface vorticity/spin line. $\bm{\sigma}^{\prime}$ is only determined by the surface pressure gradient which is solely contributed by $\bm{\sigma}_{\bm{s}}^{\prime}$. As a part of $\bm{\sigma}_{\bm{s}}^{\prime}$, the Lyman streaming vorticity flux ($\bm{\sigma}_{h}^{\prime}$) is equal to the boundary streaming vorticity flux ($\bm{\sigma}_{h}$). Correspondingly, the Lyman enstrophy dynamics is investigated, where the general formulae of the Lyman $\Omega_{R{s}}$ flux ($F_{Rs}^{\prime}$), the Lyman $\Omega_{s}$ flux ($F_{s}^{\prime}$) and the Lyman enstrophy flux ($F_{\Omega}^{\prime}$) are presented. 
Unlike the BEF $F_{\Omega}$, the quadratic form $\nu\bm{s}_{w}\bm{\cdot}\bm{K}\bm{\cdot}\bm{s}_{w}$ certainly does not contribute to both $F_{Rs}^{\prime}$ and $F_{\Omega}^{\prime}$, and  $F_{\Omega}^{\prime}$ is completely determined by the longitudinal-transverse coupling mechanism (namely, $\bm{\xi}\bm{\cdot}\bm{\nabla}_{\pi}\hat{p}_{w}$).
Overall, the Lyman vorticity and enstrophy fluxes offer a simpler physical interpretation to the vorticity and enstrophy creation on a boundary.

These derived formulae could be valuable in identifying the boundary sources of orbital rotation and spin in complex separated flows and turbulence, which could deepen the understanding of the formation of near-wall coherent structures and provide further guidance for optical configuration design and flow-noise control. In my future works, the decompositions of the boundary fluxes of vorticity and enstrophy will be generalized to an arbitrarily moving and continuously deforming wall, as well as the interface separating two immiscible bulk fluids.

\section*{CRediT authorship contribution statement} 
\textbf{Tao Chen}: Conceptualization, Methodology, Validation, Investigation, Writing --
original draft, Writing - review \& editing, Funding acquisition.
Tao Chen is responsible for the validity of the theoretical results in this paper.

\section*{Declaration of competing interest} 
The authors declare that they have no known competing financial
interests or personal relationships that could have appeared to influence
the work reported in this paper.

\section*{Data availability} 
No data is generated for the present study. 

\section*{Acknowledgements}
The work was supported by the National Natural Science Foundation of China (NSFC Award Nos. 12402262).

\appendix
\setcounter{figure}{0}
\setcounter{table}{0}
\section{Derivation of Eq.~\eqref{m2}}\label{AppendixA}
The wall-normal derivative of the unit tangent vector of a streamline $\hat{\bm{t}}=\bm{u}/q$ is calculated as
\begin{eqnarray}
\left[\partial_{n}\hat{\bm{t}}\right]_{w}\equiv\lim_{y\rightarrow0}\partial_{n}\hat{\bm{t}}=\lim_{y\rightarrow0}\frac{q\partial_{n}\bm{u}-\bm{u}\partial_{n}q}{q^2}.
\end{eqnarray}

For convenience, Eqs.~\eqref{Taylor_expansion} and~\eqref{aa1} can be simply written as
\begin{eqnarray}
	\bm{u}=\bm{a}_{1}y+\bm{a}_{2}y^2+\bm{\mathcal{O}}(y^3),
\end{eqnarray}
\begin{eqnarray}
	q=b_{1}y+b_{2}y^2+\mathcal{O}(y^3),
\end{eqnarray}
where the expansion coefficients are
\begin{equation}\label{cc1}
\bm{a}_{1}=\bm{\xi},~~\bm{a}_{2}=\frac{1}{2\nu}\bm{\nabla}_{\pi}\hat{p}_{w}+\frac{1}{2}K\bm{\xi}-\frac{1}{2}\left(\bm{\nabla}_{\pi}\bm{\cdot}\bm{\xi}\right)\bm{n},
\end{equation}
\begin{equation}\label{cc2}
b_{1}=\lVert\bm{\xi}\rVert,~~b_{2}=\frac{1}{2\nu}\partial_{\xi}\hat{p}_{w}+\frac{1}{2}K\lVert\bm{\xi}\rVert.
\end{equation}

Therefore, it follows that
\begin{eqnarray}\label{cc3}
\left[\partial_{n}\hat{\bm{t}}\right]_{w}&=&\lim_{y\rightarrow0}\frac{\left[b_{1}y+b_{2}y^2+\mathcal{O}(y^3)\right]\left[\bm{a}_{1}+2\bm{a}_{2}y+\bm{\mathcal{O}}(y^2)\right]-\left[\bm{a}_{1}y+\bm{a}_{2}y^2+\bm{\mathcal{O}}(y^3)\right]\left[b_{1}+2b_{2}y+\mathcal{O}(y^2)\right]}{\left[b_{1}y+b_{2}y^2+\mathcal{O}(y^3)\right]^2}\nonumber\\
&=&\lim_{y\rightarrow0}\frac{\left[\bm{a}_{1}b_{1}y+2\bm{a}_{2}b_{1}y^2+\bm{a}_{1}b_{2}y^{2}+\bm{\mathcal{O}}(y^3)\right]-\left[\bm{a}_{1}b_{1}y+2\bm{a}_{1}b_{2}y^2+\bm{a}_{2}b_{1}y^{2}+\bm{\mathcal{O}}(y^3)\right]}{b_1^2y^2+\mathcal{O}(y^3)}\nonumber\\
&=&\lim_{y\rightarrow0}\frac{\left[\bm{a}_{2}b_{1}y^2+\bm{\mathcal{O}}(y^3)\right]-\left[\bm{a}_{1}b_{2}y^2+\bm{\mathcal{O}}(y^3)\right]}{b_1^2y^2+\mathcal{O}(y^3)}\nonumber\\
&=&\lim_{y\rightarrow0}\frac{\bm{a}_{2}b_{1}-\bm{a}_{1}b_{2}+\bm{\mathcal{O}}(y)}{b_{1}^{2}+\bm{\mathcal{O}}(y)}\nonumber\\
&=&\frac{\bm{a}_{2}b_{1}-\bm{a}_{1}b_{2}}{b_{1}^{2}}.
\end{eqnarray}

Combining Eqs.~\eqref{cc1},~\eqref{cc2} and~\eqref{cc3} yields
\begin{eqnarray}
\left[\partial_{n}\hat{\bm{t}}\right]_{w}=\frac{1}{\lVert\bm{\xi}\rVert}\left[\frac{1}{2\nu}\bm{e}_{\omega}\partial_{\omega}\hat{p}_{w}-\frac{1}{2}\left(\bm{\nabla}_{\pi}\bm{\cdot}\bm{\xi}\right)\bm{n}\right].
\end{eqnarray}

\section{Derivation of Eq.~\eqref{BFI}}\label{new81}
The boundary $\bm{\omega}_{\rm{I}}$ flux [for $\bm{\omega}_{\rm{I}}$ in Eq.~\eqref{omega1}] is evaluated as
\begin{eqnarray}\label{ff1}
	\bm{\sigma}_{\rm{I}}
	&=&\nu\left[\partial_{n}{q}\right]_{w}\left[\bm{\nabla}\times\hat{\bm{t}}\right]_{w}
	+\nu q_{w}\left[\partial_{n}\left(\bm{\nabla}\times\hat{\bm{t}}\right)\right]_{w}\nonumber\\
	&=&\nu\left[\partial_{n}{q}\right]_{w}\left[\bm{\nabla}\times\hat{\bm{t}}\right]_{w}.
\end{eqnarray}

The curl of the unit tangent vector $\hat{\bm{t}}$ of a streamline  can be decomposed as
\begin{eqnarray}\label{eq15}
	\left[\bm{\nabla}\times\hat{\bm{t}}\right]_{w}=\bm{\nabla}_{\pi}\times\hat{\bm{t}}_{w}+\bm{n}\times\left[\partial_{n}\hat{\bm{t}}\right]_{w}.
\end{eqnarray} 
In the right hand side of Eq.~\eqref{eq15}, the first term is evaluated as
\begin{eqnarray}\label{eq16}
	\bm{\nabla}_{\pi}\times\hat{\bm{t}}_{w}=\frac{1}{\lVert\bm{\xi}\rVert}\bm{\nabla}_{\pi}\times\bm{\xi}-\frac{1}{\lVert\bm{\xi}\rVert^{2}}\bm{\nabla}_{\pi}\lVert\bm{\xi}\rVert\times\bm{\xi}.
\end{eqnarray}
One one hand, it holds that
\begin{eqnarray}\label{eq17}
	\bm{\nabla}_{\pi}\times\bm{\xi}&=&\bm{\nabla}_{\pi}\times\left(\bm{s}_{w}\times\bm{n}\right)\nonumber\\
	&=&\bm{n}\bm{\cdot}\bm{\nabla}_{\pi}\bm{s}_{w}-\bm{s}_{w}\bm{\cdot}\bm{\nabla}_{\pi}\bm{n}
	+\left(\bm{\nabla}_{\pi}\bm{\cdot}\bm{n}\right)\bm{s}_{w}-\left(\bm{\nabla}_{\pi}\bm{\cdot}\bm{s}_{w}\right)\bm{n}\nonumber\\
	&=&\bm{K}\bm{\cdot}\bm{s}_{w}-K\bm{s}_{w}-\left(\bm{\nabla}_{\pi}\bm{\cdot}\bm{s}_{w}\right)\bm{n}.
\end{eqnarray}
On the other hand, we have
\begin{eqnarray}\label{eq18}
	-\frac{1}{\lVert\bm{\xi}\rVert^{2}}\bm{\nabla}_{\pi}\lVert\bm{\xi}\rVert\times\bm{\xi}
	=\frac{1}{\lVert\bm{\xi}\rVert}\bm{e}_{\xi}\times\bm{\nabla}_{\pi}\lVert\bm{\xi}\rVert
	=\frac{1}{\lVert\bm{\xi}\rVert}\partial_{\omega}\lVert\bm{s}_{w}\rVert\bm{n}.
\end{eqnarray}
By using Eqs.~\eqref{eq17} and~\eqref{eq18}, Eq.~\eqref{eq16} becomes
\begin{eqnarray}\label{eq19}
	\bm{\nabla}_{\pi}\times\hat{\bm{t}}_{w}=\frac{1}{\lVert\bm{\xi}\rVert}
	\left[\bm{K}\bm{\cdot}\bm{s}_{w}-K\bm{s}_{w}-\left(\bm{\nabla}_{\pi}\bm{\cdot}\bm{s}_{w}\right)\bm{n}+\partial_{\omega}\lVert\bm{s}_{w}\rVert\bm{n}\right].
\end{eqnarray}
By Eq.~\eqref{m2}, the second term in the right hand side of Eq.~\eqref{eq15} is calculated as
\begin{eqnarray}\label{eq20}
	\bm{n}\times\left[\partial_{n}\hat{\bm{t}}\right]_{w}=-\frac{1}{\lVert\bm{\xi}\rVert}\frac{1}{2\nu}\bm{e}_{\xi}\partial_{\omega}\hat{p}_{w}.
\end{eqnarray}
Combining Eqs.~\eqref{eq15},~\eqref{eq19} and~\eqref{eq20} ends the proof of Eq.~\eqref{BFI}.

\section{Derivation of Eq.~\eqref{BFII}}\label{new82}
The boundary $\bm{\omega}_{\rm{II}}$ flux [for $\bm{\omega}_{\rm{II}}$ in Eq.~\eqref{omega2}] is evaluated as
\begin{eqnarray}\label{BSF}
	\bm{\sigma}_{\rm{II}}\equiv\nu\left[\partial_{n}\left(\bm{\nabla}q\times\hat{\bm{t}}\right)\right]_{w}
	=\nu\left[\partial_{n}\bm{\nabla}q\right]_{w}\times\hat{\bm{t}}_{w}
	+\nu\left[\bm{\nabla}q\right]_{w}\times\left[\partial_{n}\hat{\bm{t}}\right]_{w}.
\end{eqnarray}
In the right hand side of Eq.~\eqref{BSF}, the first term is 
\begin{eqnarray}\label{BSF_term1}
	\nu\left[\partial_{n}\bm{\nabla}q\right]_{w}\times\hat{\bm{t}}_{w}=\nu\left\{\bm{\nabla}_{\pi}\left[\partial_{n}q\right]_{w}+\bm{n}\left[\partial_{n}^{2}q\right]_{w}\right\}\times\bm{e}_{\xi}
\end{eqnarray}
Then, substituting Eqs.~\eqref{m3} and~\eqref{m4} into Eq.~\eqref{BSF_term1} gives
\begin{eqnarray}\label{BSF_term1a}
	\nu\left[\partial_{n}\bm{\nabla}q\right]_{w}\times\hat{\bm{t}}_{w}=-\nu\partial_{\omega}\lVert\bm{s}_{w}\rVert\bm{n}+\bm{e}_{\omega}\partial_{\xi}\hat{p}_{w}+\nu{K}\bm{s}_{w}.
\end{eqnarray}
By Eqs.~\eqref{m2} and~\eqref{m3}, the second term becomes
\begin{eqnarray}\label{BSF_term2}
	\nu\left[\bm{\nabla}q\right]_{w}\times\left[\partial_{n}\hat{\bm{t}}\right]_{w}
	=-\frac{1}{2}\bm{e}_{\xi}\partial_{\omega}\hat{p}_{w}.
\end{eqnarray}

Combining Eqs.~\eqref{BSF},~\eqref{BSF_term1a} and~\eqref{BSF_term2} yields
\begin{eqnarray}\label{BSF_f11}
	\bm{\sigma}_{\rm{II}}
	=-\nu\partial_{\omega}\lVert\bm{s}_{w}\rVert\bm{n}
	+\bm{e}_{\omega}\partial_{\xi}\hat{p}_{w}
	-\frac{1}{2}\bm{e}_{\xi}\partial_{\omega}\hat{p}_{w}
	+\nu{K}\bm{s}_{w}.
\end{eqnarray}
Then, applying the relation $\bm{n}\times\bm{\nabla}_{\pi}\hat{p}_{w}=\bm{e}_{\omega}\partial_{\xi}\hat{p}_{w}-\bm{e}_{\xi}\partial_{\omega}\hat{p}_{w}$
yields Eq.~\eqref{BFII}.

\section{Compact representation of several boundary vorticity fluxes}\label{BVFsCR}
Interestingly, by virtue of the wedge product and the Hodge star operator~\cite{ChenWH2002}, we find that
\begin{subequations}
	\begin{eqnarray}
		K\bm{s}_{w}+\nu^{-1}\bm{n}\times\bm{\nabla}_{\pi}\hat{p}_{w}=*\left[\bm{n}\wedge\left(K\bm{\xi}+\nu^{-1}\bm{\nabla}_{\pi}\hat{p}_{w}\right)\right],
	\end{eqnarray}
	\begin{eqnarray}
		\bm{K}\bm{\cdot}\bm{s}_{w}-K\bm{s}_{w}-\left(\bm{\nabla}_{\pi}\bm{\cdot}\bm{s}_{w}\right)\bm{n}=*\left(\bm{\nabla}_{\pi}\wedge\bm{\xi}\right).
	\end{eqnarray}
\end{subequations}

Then, the boundary fluxes of $\bm{\omega}_{\rm{I}}$ and $\bm{\omega}_{\rm{II}}$ in Eqs.~\eqref{BFI} and~\eqref{BFII} can be expressed in equivalent but more compact forms:
\begin{subequations}
	\begin{eqnarray}\label{BSF_f1}
		\bm{\sigma}_{\rm{I}}=-\nu{\partial_{\omega}\lVert\bm{s}_{w}\rVert}\bm{n}
		+\frac{1}{2}\bm{e}_{\xi}{\partial_{\omega}{\hat{p}_{w}}}+\nu*\left[\bm{n}\wedge\left(K\bm{\xi}+\nu^{-1}\bm{\nabla}_{\pi}\hat{p}_{w}\right)\right],
	\end{eqnarray}
	\begin{eqnarray}\label{BRF1}
		\bm{\sigma}_{\rm{II}}=\nu{\partial_{\omega}\lVert\bm{s}_{w}\rVert}\bm{n}-\frac{1}{2}\bm{e}_{\xi}{\partial_{\omega}{\hat{p}_{w}}}+\nu*\left(\bm{\nabla}_{\pi}\wedge\bm{\xi}\right).
	\end{eqnarray}
\end{subequations}
Consequently, the total BVF $\bm{\sigma}$ can be viewed as the dual vector of a $2$-form $\bm{\Phi}$: 
\begin{eqnarray}
	\bm{\sigma}=*\bm{\Phi},~~~~\bm{\Phi}\equiv\nu\bm{n}\wedge\left(K\bm{\xi}+\nu^{-1}\bm{\nabla}_{\pi}\hat{p}_{w}\right)+\nu\bm{\nabla}_{\pi}\wedge\bm{\xi}.
\end{eqnarray}

\section{Orientation-dependent intrinsic expression of orbital rotation}\label{AppendixB}
The orbital rotation $\bm{\psi}$ can be written as
\begin{eqnarray}\label{ct1}
	\bm{\psi}=\kappa{q}\hat{\bm{b}}=\kappa{q}\hat{\bm{t}}\times\hat{\bm{n}}=\left(q\hat{\bm{t}}\right)\times\left(\kappa\hat{\bm{n}}\right)={\bm{u}}\times\partial_{s}\hat{\bm{t}}.
\end{eqnarray}

The derivative of $\hat{\bm{t}}$ with respect to the arc length parameter $s$ of a streamline is evaluated as
\begin{eqnarray}\label{ct2}
	\partial_{s}\hat{\bm{t}}=\partial_{s}\left(\frac{\bm{u}}{q}\right)
	=\frac{\left(\partial_{s}\bm{u}\right)q-\bm{u}\left(\partial_{s}q\right)}{q^{2}}.
\end{eqnarray}

Since $\partial_{s}=\hat{\bm{t}}\bm{\cdot}\bm{\nabla}$, Eq.~\eqref{ct2} can be transformed as
\begin{eqnarray}\label{ct3}
	\partial_{s}\hat{\bm{t}}&=&\frac{\left(\hat{\bm{t}}\bm{\cdot}\bm{\nabla}\bm{u}\right)q-\bm{u}\left(\hat{\bm{t}}\bm{\cdot}\bm{\nabla}q\right)}{q^{2}}\nonumber\\
	&=&\frac{\bm{u}\bm{\cdot}\bm{\nabla}\bm{u}}{q^2}-\hat{\bm{t}}\left(\hat{\bm{t}}\bm{\cdot}\frac{1}{q}\bm{\nabla}q\right).
\end{eqnarray}

By acting the gradient operator to both side of $q^{2}=\bm{u}\bm{\cdot}\bm{u}$, one can obtain
\begin{eqnarray}\label{ct4}
	\frac{1}{q}\bm{\nabla}q=\frac{1}{q^{2}}\bm{\nabla}\bm{u}\bm{\cdot}\bm{u}.
\end{eqnarray}
Therefore, substituting Eq.~\eqref{ct4} into Eq.~\eqref{ct3} yields
\begin{eqnarray}\label{ct5}
	\partial_{s}\hat{\bm{t}}&=&\frac{\bm{u}\bm{\cdot}\bm{\nabla}\bm{u}}{q^2}-\frac{1}{q^{2}}\hat{\bm{t}}\left(\hat{\bm{t}}\bm{\cdot}\bm{\nabla}\bm{u}\bm{\cdot}\bm{u}\right)\nonumber\\
	&=&\frac{1}{q^{2}}\left[\bm{u}\bm{\cdot}\bm{\nabla}\bm{u}-q\hat{\bm{t}}\left(\hat{\bm{t}}\bm{\cdot}\bm{\nabla}\bm{u}\bm{\cdot}\hat{\bm{t}}\right)\right]\nonumber\\
	&=&\frac{1}{q^{2}}\left[\bm{u}\bm{\cdot}\bm{\nabla}\bm{u}-q\hat{\bm{t}}\left(\hat{\bm{t}}\bm{\cdot}\bm{\nabla}\bm{u}^{T}\bm{\cdot}\hat{\bm{t}}\right)\right]\nonumber\\
	&=&\frac{1}{q^{2}}\left[\bm{u}\bm{\cdot}\bm{\nabla}\bm{u}-\hat{\bm{t}}\left(\hat{\bm{t}}\bm{\cdot}\bm{\nabla}\bm{u}^{T}\bm{\cdot}\bm{u}\right)\right]\nonumber\\
	&=&\frac{1}{q^{2}}\left[\bm{u}\bm{\cdot}\bm{\nabla}\bm{u}-\hat{\bm{t}}\hat{\bm{t}}\bm{\cdot}\left(\bm{u}\bm{\cdot}\bm{\nabla}\bm{u}\right)\right]\nonumber\\
	&=&-\frac{\hat{\bm{t}}\times\left[\hat{\bm{t}}\times\left(\bm{u}\bm{\cdot}\bm{\nabla}\bm{u}\right)\right]}{q^{2}}.
\end{eqnarray}

From Eqs.~\eqref{ct1} and~\eqref{ct5}, the intrinsic expression of the orbital rotation is derived as
\begin{eqnarray}\label{ct6}
	\bm{\psi}=\frac{1}{q^{2}}\bm{u}\times\left(\bm{u}\bm{\cdot}\bm{\nabla}\bm{u}\right)=\hat{\bm{t}}\times\left(\hat{\bm{t}}\bm{\cdot}\bm{\nabla}\bm{u}\right)=\hat{\bm{t}}\times\left(\hat{\bm{t}}\bm{\cdot}\bm{A}\right).
\end{eqnarray}

\section{Derivation of Eq.~\eqref{sigma_psi}}\label{boundary_OR_flux_derivation}
The boundary orbital-rotation flux in Eq.~\eqref{BVF_LPWu} is evaluated as
\begin{eqnarray}\label{BOF2}
	\bm{\sigma}_{\bm{\psi}}=\nu\left[\partial_{n}\hat{\bm{t}}\right]_{w}\times\left(\hat{\bm{t}}_{w}\bm{\cdot}\bm{A}_{w}\right)+\nu\hat{\bm{t}}_{w}\times\left(\left[\partial_{n}\hat{\bm{t}}\right]_{w}\bm{\cdot}\bm{A}_{w}\right)
	+\nu\hat{\bm{t}}_{w}\times\left(\hat{\bm{t}}_{w}\bm{\cdot}\left[\partial_{n}\bm{A}\right]_{w}\right).
\end{eqnarray}
In the right hand side of Eq.~\eqref{BOF2}, by using Eq.~\eqref{m1} and $\bm{A}_{w}=\bm{n}\bm{\xi}$, the first term vanishes because
\begin{eqnarray}\label{BOF3}
	\hat{\bm{t}}_{w}\bm{\cdot}\bm{A}_{w}=\bm{e}_{\xi}\bm{\cdot}\bm{n}\bm{\xi}=\bm{0}.
\end{eqnarray}
By applying Eqs.~\eqref{m1} and~\eqref{m2}, the second term is evaluated as
\begin{eqnarray}\label{BOF4}
	\nu\hat{\bm{t}}_{w}\times\left(\left[\partial_{n}\hat{\bm{t}}\right]_{w}\bm{\cdot}\bm{A}_{w}\right)
	=-\frac{1}{2}\left(\bm{\nabla}_{\pi}\bm{\cdot}\bm{\xi}\right)\bm{e}_{\xi}\times\bm{e}_{\xi}=\bm{0}.
\end{eqnarray}
Since the boundary $\bm{A}$ flux is expressed as~\cite{ChenWu2024}
\begin{eqnarray}\label{BOF5}
	\left[\partial_{n}\bm{A}\right]_{w}=\bm{\nabla}_{\pi}\bm{\xi}+\bm{n}\left[K\bm{\xi}+\nu^{-1}\bm{\nabla}_{\pi}\hat{p}_{w}-(\bm{\nabla}_{\pi}\bm{\cdot}\bm{\xi})\bm{n}\right],
\end{eqnarray}
the last term is calculated as
\begin{eqnarray}\label{BOF6}
	\nu\hat{\bm{t}}_{w}\times\left(\hat{\bm{t}}_{w}\bm{\cdot}\left[\partial_{n}\bm{A}\right]_{w}\right)
	=\nu\bm{e}_{\xi}\times\left(\bm{e}_{\xi}\bm{\cdot}\bm{\nabla}_{\pi}\bm{\xi}\right).
\end{eqnarray}
Combining Eqs.~\eqref{BOF2}--~\eqref{BOF6} gives $\bm{\sigma}_{\bm\psi}$ in Eq.~\eqref{sigma_psi}. 

To evaluate the components of $\bm{\sigma}_{\bm\psi}$, Eq.~\eqref{sigma_psi} can be written as
\begin{eqnarray}\label{qqq1}
\bm{\sigma}_{\bm\psi}=\nu\left(\bm{e}_{\xi}\bm{\cdot}\bm{\nabla}_{\pi}\bm{\xi}\bm{\cdot}\bm{e}_{\omega}\right)\bm{n}-\nu\left(\bm{e}_{\xi}\bm{\cdot}\bm{\nabla}_{\pi}\bm{\xi}\bm{\cdot}\bm{n}\right)\bm{e}_{\omega}.
\end{eqnarray}
The term in the first bracket of Eq.~\eqref{qqq1} is evaluated as
\begin{eqnarray}\label{qqq2}
\bm{e}_{\xi}\bm{\cdot}\bm{\nabla}_{\pi}\bm{\xi}\bm{\cdot}\bm{e}_{\omega}=-\frac{1}{2\Omega_{w}}\bm{\xi}\bm{\cdot}\bm{\nabla}_{\pi}\bm{s}_{w}\bm{\cdot}\bm{\xi}.
\end{eqnarray}
By noticing $\bm{\xi}\times\bm{s}_{w}=2\Omega_{w}\bm{n}$ and the vector identity
\begin{eqnarray}\label{qqq3}
\bm{\nabla}_{\pi}\times\left(\bm{\xi}\times\bm{s}_{w}\right)
=\bm{s}_{w}\bm{\cdot}\bm{\nabla}_{\pi}\bm{\xi}-\bm{\xi}\bm{\cdot}\bm{\nabla}_{\pi}\bm{s}_{w}+\left(\bm{\nabla}_{\pi}\bm{\cdot}\bm{s}_{w}\right)\bm{\xi}
-\left(\bm{\nabla}_{\pi}\bm{\cdot}\bm{\xi}\right)\bm{s}_{w},
\end{eqnarray}
we obtain
\begin{eqnarray}\label{qqq4}
\bm{s}_{w}\bm{\cdot}\bm{\nabla}_{\pi}\Omega_{w}
-2\left(\bm{\nabla}_{\pi}\bm{\cdot}\bm{s}_{w}\right)\Omega_{w}=-\bm{\xi}\bm{\cdot}\bm{\nabla}_{\pi}\bm{s}_{w}\bm{\cdot}\bm{\xi}.
\end{eqnarray}
Combining Eqs.~\eqref{qqq2} and~\eqref{qqq4} gives
\begin{eqnarray}\label{qqq5}
\bm{e}_{\xi}\bm{\cdot}\bm{\nabla}_{\pi}\bm{\xi}\bm{\cdot}\bm{e}_{\omega}
=\partial_{\omega}\lVert\bm{s}_{w}\rVert-\bm{\nabla}_{\pi}\bm{\cdot}\bm{s}_{w}.
\end{eqnarray}
The surface divergence of the spin is expanded as
\begin{eqnarray}\label{qqq6}
\bm{\nabla}_{\pi}\bm{\cdot}\bm{s}_{w}=\partial_{\omega}\lVert\bm{s}_{w}\rVert-\kappa_{g,\xi}\lVert\bm{s}_{w}\rVert.
\end{eqnarray}
where $\kappa_{g,\xi}$ is the geodesic curvature of the $\bm{\xi}$-line.
Substituting Eq.~\eqref{qqq6} into Eq.~\eqref{qqq5} yields
\begin{eqnarray}\label{qqq7}
\nu\left(\bm{e}_{\xi}\bm{\cdot}\bm{\nabla}_{\pi}\bm{\xi}\bm{\cdot}\bm{e}_{\omega}\right)=\nu\kappa_{g,\xi}\lVert\bm{s}_{w}\rVert.
\end{eqnarray}
The term in the second bracket of Eq.~\eqref{qqq1} is evaluated as
\begin{eqnarray}\label{qqq8}
\nu\left(\bm{e}_{\xi}\bm{\cdot}\bm{\nabla}_{\pi}\bm{\xi}\bm{\cdot}\bm{n}\right)
=\nu\lVert\bm{s}_{w}\rVert^{-1}\bm{\xi}\bm{\cdot}\bm{K}\bm{\cdot}\bm{\xi}=\nu{b}_{\xi\xi}\lVert\bm{s}_{w}\rVert.
\end{eqnarray}
Substituting Eqs.~\eqref{qqq7} and~\eqref{qqq8} into Eq.~\eqref{qqq1} yields Eq.~\eqref{sigma_psi}.


\bibliography{mybibfile}

\begin{thebibliography}{10}
\expandafter\ifx\csname url\endcsname\relax
  \def\url#1{\texttt{#1}}\fi
\expandafter\ifx\csname urlprefix\endcsname\relax\def\urlprefix{URL }\fi
\expandafter\ifx\csname href\endcsname\relax
  \def\href#1#2{#2} \def\path#1{#1}\fi

\bibitem{Batchelor1967}
G.~K. Batchelor, An introduction to fluid dynamics, Cambridge University Press,
  Cambridge, 1967.

\bibitem{WuJZ2015book}
J.-Z. Wu, H.-Y. Ma, M.-D. Zhou, {V}ortical {F}lows, Springer, Berlin,
  Heidelberg, 2015.

\bibitem{Lighthill1963}
M.~J. Lighthill, {Introduction of Boundary layer Theory}, in: L.~Rosenhead
  (Ed.), Laminar boundary layers, Vol.~I, Oxford University Press, Oxford,
  1963, pp. 46--113.

\bibitem{Panton1984}
R.~L. Panton, Incompressible flows, John Wiley \& Sons, United States, 1984.

\bibitem{WuJZWuJM1993}
J.~Z. Wu, J.~M. Wu, Interactions between a solid surface and a viscous
  compressible flow field, J. Fluid Mech. 254 (1993) 183--211.

\bibitem{Lyman1990}
F.~A. Lyman, Vorticity production at a solid boundary, Appl. Mech. Rev. 43~(8)
  (1990) 157--158,
  https://www.cfm.brown.edu/faculty/gk/AM258/Handouts/258-Psi-Omega.pdf.

\bibitem{Terrington2023JFM}
S.~J. Terrington, K.~Hourigan, M.~C. Thompson, The {L}yman-{H}uggins
  interpretation of enstrophy transport, J. Fluid Mech. 958 (2023) A30.

\bibitem{Chen2024AMS}
T.~Chen, C.~Wang, T.~Liu, On physics of boundary vorticity creation in
  incompressible viscous flow, Acta Mech. Sin. 40 (2024) 323443.

\bibitem{XinWu2013}
Z.~Q. Xin, C.~J. Wu, Vorticity dynamics and control of self-propelled swimming
  of the 3{D} rays, in: Proceedings of the 14th Asia Congress of Fluid
  Mechanics, Hanoi and Halong, Vietnam, 2013.

\bibitem{ChenYu2017}
Y.~Chen, Theoretical and numerical studies on vorticity dynamics of flow with
  deformable boundary, Ph.D. thesis, Fudan University, Shanghai, China (2017).

\bibitem{WuJZ1995}
J.-Z. Wu, A theory of three-dimensional interfacial vorticity dynamics, Phys.
  Fluids 7 (1995) 2375.

\bibitem{Liu2016MST}
T.~Liu, T.~Misaka, K.~Asai, S.~Obayashi, J.-Z. Wu, Feasibility of skin-friction
  diagnostics based on surface pressure gradient field, Meas. Sci. Technol
  27~(12) (2016) 125304.

\bibitem{ChenTao2019POF}
T.~Chen, T.~Liu, L.-P. Wang, S.~Chen, Relations between skin friction and other
  surface quantities in viscous flows, Phys. Fluids 31~(10) (2019) 107101.

\bibitem{Liu2019PAS}
T.~Liu, Global skin friction measurements and interpretation, Prog. Aeosp. Sci.
  111 (2019) 100584.

\bibitem{ChenTao2021POF}
T.~Chen, T.~Liu, Z.-Q. Dong, L.-P. Wang, S.~Chen, Near-wall flow structures and
  related surface quantities in wall-bounded turbulence, Phys. Fluids 33~(6)
  (2021) 065116.

\bibitem{Chen2023PhysicaD}
T.~Chen, T.~Liu, Lamb dilatation and its hydrodynamic viscous flux in near-wall
  incompressible flows, Physica D: Nonlinear Phenomena 448 (2023) 133730.

\bibitem{ChenLiu2023LD}
T.~Chen, T.~Liu, Lie derivatives of fundamental surface quantities in
  incompressible viscous flows, Phys. Fluids 35~(5) (2023) 057104.

\bibitem{Terrington2022JFM}
S.~J. Terrington, K.~Hourigan, M.~C. Thompson, Vorticity generation and
  conservation on generalised interfaces in three-dimensional flows, J. Fluid
  Mech. 936 (2022) A44.

\bibitem{ChenTao2024IJMF}
T.~Chen, C.~Wang, T.~Liu, Interfacial vorticity dynamics for
  {N}avier-{S}tokes-{K}orteweg system: {G}eneral theory and application to
  two-dimensional near-wall cavitation bubble, Int. J. Multiphas. Flow 172
  (2024) 104705.

\bibitem{ChenWu2024}
T.~Chen, J.-Z. Wu, T.~Liu, D.~M. Salazar, Boundary sources of velocity gradient
  tensor and its invariants, arXiv:2406.08760 (June 2024).

\bibitem{Serrin1959}
J.~Serrin, Mathematical Principles of Classical Fluid Mechanics, Springer
  Berlin Heidelberg, Berlin, Heidelberg, 1959, pp. 125--263.

\bibitem{WuChen2024}
T.~Chen, A.-K. Gao, J.-Z. Wu, Fluid-element rotation and vorticity splitting,
  (unpublished) (August 2024).

\bibitem{LiuCQ2018}
C.~Liu, Y.~Gao, S.~Tian, X.~Dong, Rortex—a new vortex vector definition and
  vorticity tensor and vector decompositions, Phys. Fluids 30 (2018) 035103.

\bibitem{LiZhen2024}
Z.~Li, Schur {F}orms and {N}ormal-{N}ilpotent {D}ecompositions, Applied
  Mathematics and Mechanics 45~(9) (2024) 1--12.

\bibitem{ChenWH2002}
W.~H. Chen, X.~X. Li, An introduction to Riemann geometry, Peking University
  Press, Beijing, 2002, (in Chinese).

\end{thebibliography}

\end{document}